\documentstyle[prl,aps,epsf,float,graphicx]{revtex}
\DeclareGraphicsExtensions{.eps,.ps}
\begin{document}
\draft
\twocolumn[\hsize\textwidth\columnwidth\hsize\csname@twocolumnfalse\endcsname
\title{Electronic structure of multiquantum giant vortex states
in mesoscopic superconducting disks}
\author{K. Tanaka$^{1,2}$, Istv\'an Robel$^{1,3}$, and
Boldizs\'ar Jank\'o$^{1,3}$ }
\address{$^1$Materials Science Division, Argonne National Laboratory,\\
9700 South Cass Avenue, Argonne, IL 60439}
\address{$^2$Department of Physics and Engineering Physics,
University of Saskatchewan,\\ Saskatoon, SK, Canada S7N 5E2}
\address{$^3$Department of Physics,
University of Notre Dame, Notre Dame, IN 46556-5670}
\date{Published in {\it Proc. Natl. Acad. Sci.} USA {\bf 99}, 5233-5236 (2002).}
\maketitle

\begin{abstract}

  We report self-consistent calculations of the microscopic electronic
  structure of the so-called giant vortex states. These novel
  multiquantum vortex states, detected by recent magnetization measurements
  on submicron disks, are qualitatively different from the Abrikosov vortices
  in the bulk. We find that, in addition to multiple branches of bound
  states in the core region, the local tunneling density of states exhibits
  Tomasch oscillations due to the single-particle interference arising from
  quantum confinement. These features should be directly observable by
  scanning tunneling spectroscopy.

\end{abstract}
 ]

\makeatletter
\global\@specialpagefalse

\let\@evenhead\@oddhead
\makeatother

\narrowtext

\section{Introduction}

\label{sec:intro} Superconducting vortices are topological
singularities in the order parameter \cite{abrikosov}.  In a bulk
system each vortex carries a single flux quantum, while vortices
with multiple flux quanta are not favorable energetically
\cite{fetter}. In small superconductors, however, the situation
may be different.  Today's nanotechnology can provide valuable
insight into the nature of mesoscopic superconductors, whose
linear dimensions can be comparable to the coherence length or the
inter-vortex distance of the Abrikosov lattice. The following
question then arises naturally: does single-quantum vortex matter
survive the limit of decreasing sample size?  More than thirty
years ago, Fink and Presson gave the intriguing answer ``not
always'' in their pioneering work \cite{fink66} on a related
system: a thin cylinder in parallel field. They have shown it
theoretically, within the framework of the phenomenological
Ginzburg-Landau (GL) theory, and also provided experimental
evidence \cite{finkpl} for the existence of an enormous superfluid
eddy current on the surface of a thin cylinder. They called this
state a giant vortex state. While the work of Fink and Presson was
largely forgotten for the next several decades, it nevertheless
anticipated the present excitement in the field of nanoscale
superconductivity.  With recent advances in the controlled
fabrication and study of nanometer-scale superconductors, the
concept of giant vortex has been brought back to focus by
Moshchalkov and coworkers a few years ago \cite{moshchalkov95}.
Their experiments on mesoscopic squares and square rings have
indeed revealed that small superconductors do not always favor
many-vortex states reminiscent of the Abrikosov vortex lattice.
The measured {\it H-T} phase boundaries of these small structures
were explained in terms of giant vortex states in the GL picture
\cite{moshchalkov95}. Subsequent experiments on submicron disks
\cite{geim97} have further shown the existence of giant vortex
states inside the phase boundaries. Within the GL framework
\cite{geim97,deo97,schweigert98,deo99,schweigert99} some of the
abrupt changes in the magnetization observed have been attributed,
e.g., to the collapse of a multi-vortex state into a giant vortex,
or to transitions among different giant vortex states.

This new phase of vortex matter has a single vortex occupying the
sample, carrying multiple fluxoid \cite{fink66,degennes66,geim00}
quanta. Such a state has no immediate analogue in bulk systems,
and would only be similar to vortex states predicted for
artificially patterned structures \cite{tanaka93,columnar}.
Moshchalkov {\it et al.} \cite{moshchalkov97} have also suggested
that giant vortex states can cause the peculiar paramagnetic
Meissner effect seen in granular and mesoscopic superconductors
\cite{paramag1,paramag2}, through the compression of the flux
trapped in the sample. This effect has been seen experimentally in
mesoscopic systems by Geim and coworkers \cite{geim98}.  As it is
apparent from these recent experiments, mesoscopic superconductors
exhibit novel quantum phenomena that are not observable in bulk
systems.  Studying their unique properties is crucial not only for
potential applications but also for better understanding of
nanoscale superconductivity.

Despite the fact that the existence of giant vortex states has
been indicated more than three decades ago, and that their
counterpart in nanoscale superconductors has been under intense
scrutiny in recent years, there has been no microscopic,
self-consistent theoretical study of its electronic structure.  In
this paper we report the results of such a microscopic and
self-consistent study of multiquantum giant vortex states in
$s$-wave superconducting disks of submicron size,\footnote{After
the completion of this work, we became aware of the recent results
of A. S. Mel'nikov and V. M. Vinokur \cite{melnikov02}, in which
the possible application of mesoscopic disks as quantum switches
is studied. This work discusses the tunneling density of states in
the core of giant vortices in terms of quasiclassical
calculations.} using the Bogoliubov-de Gennes (BdG) formalism
\cite{degennes66}. The spectroscopic properties we have obtained
for such vortex states can be probed directly by scanning
tunneling microscopy (STM). Although GL studies give a good
qualitative picture for a wide range of parameters, quantitatively
reliable results are limited to the range relatively close to the
critical temperature and magnetic field. Furthermore, analyzing
the system from a microscopic point of view is essential to
understanding superconductivity on such a small scale. Latest
experimental efforts aimed at STM imaging of mesoscopic vortex
matter have focused on ${\rm NbSe}_2$ samples \cite{Wai}, since
direct STM images of vortex states have been obtained only on
high-quality single crystals of ${\rm NbSe}_2$ \cite{hess89} and
${\rm Bi_2Sr_2CaCuO_{8+\delta}}$ \cite{renner,kugler,pan}.  These
compounds are highly two-dimensional and easy to cleave {\it in
situ}, providing very clean surfaces -- a key ingredient for
successful STM imaging. Thus, anticipating STM measurements, we
present in this paper self-consistent BdG results with parameters
corresponding to submicron ${\rm NbSe}_2$ disks.

\section{Formulation}
\label{sec:form}
We consider an $s$-wave superconducting disk of radius $R$ under a
magnetic field perpendicular to the disk area, with a vortex
carrying $m$ fluxoid quanta formed in the center. The system has
cylindrical symmetry and it is described using cylindrical
coordinates $(r,\theta ,z)$. In accordance with the experiments
\cite{geim97,Wai} we assume the disk thickness to be much smaller
than the penetration depth. Consequently, the order parameter is
assumed to be uniform in the field direction ($\hat z$), and the
current density ${\bf j}$ as well as the vector potential ${\bf
A}$ has no $z$ component \cite{deo97}. In the gauge which removes
the phase of the order parameter \cite{degennes66}, i.e., $\Delta
({\bf r})=|\Delta (r)|$, we can write down the radial part of the
BdG equations \cite{gygi91,tanaka93,virtanen99} as
\begin{eqnarray}
\sigma _z {\frac{\hbar ^2}{2m_e}}\biggl[ -\biggl( {\frac{d^2}{dr^2}}+{%
\frac 1r}{\frac d{dr}}\biggr) \biggr.  \nonumber \\ {} + \biggl.
{\frac 1{r^2}} \biggl\{ l-\sigma _z \biggl(  {\frac e{\hbar c}}
r\tilde A_\theta (r)+{\frac m2}\biggr) \biggr\} ^2 \biggr] \hat
\psi _n(r) \nonumber \\ {} + \sigma_z(U(r)-\mu )\hat \psi _n(r)
 {} + \sigma _x |\Delta (r)| \hat
\psi _n(r)=\epsilon_n \hat \psi _n(r), \label{bdgeq2}
\end{eqnarray}
where
\begin{equation}
\hat \psi _n(r)=\biggl(
\begin{array}{c}
u_n(r) \\ v_n(r)
\end{array}
\biggr) \label{bdgwf2}
\end{equation}
is the radial quasiparticle amplitude. Here $\sigma _x$ and
$\sigma _z$ are the Pauli matrices, $m_e$ is the electron mass,
and $\mu $ the chemical potential. The angular momentum quantum
number $l$ is an integer when $m$ is even, and a half odd integer
when $m$ is odd \cite{degennes66}. The single-particle potential
$U(r)$ can incorporate the lattice potential and inhomogeneity
effects due to impurities and sample boundaries. To study quantum
size and interference effects, we consider a clean sample and take
$U(r)=0$ inside the disk, while including the periodic lattice
potential in terms of the effective masses, $m_r$ and $m_z$.
Furthermore, we take $m_z\gg m_r$, as justified for highly
anisotropic materials such as ${\rm NbSe}_2$, and neglect the
dependence of Eq.~(\ref{bdgeq2}) on the motion along the $z$
direction.

In principle, due to finite thickness of the sample, the vector
potential must depend on not only $r$ but also $z$, so that ${\bf
A}=A_\theta(r,z) \hat\theta$, and $\tilde A_{\theta}(r)$ in
(\ref{bdgeq2}) is an average of $A_\theta(r,z)$ over the disk
height $L$ \cite{deo97}:
\begin{equation}
\tilde A_{\theta}(r) = {\frac{1}{L}}\int_{-L/2}^{L/2}dz
\,A_\theta(r,z). \label{aint}
\end{equation}
However, for typical experimental parameters \cite{geim97,Wai} the
lateral sample size is much larger than the thickness, and we
therefore consider the case where the vector potential is independent of the $%
z$ coordinate; $A_\theta(r,z) \simeq A_{\theta}(r)$. The order
parameter and the current density ${\bf j}\equiv j_\theta(r)
\hat\theta$ are given in terms of the eigenvalues and
eigenfunctions of Eq.~(\ref{bdgeq2}) as
\begin{eqnarray}
|\Delta(r)| = g\sum_{\epsilon_n\le \hbar\omega_D} u_n(r)
v_n^\ast(r) (1 - 2f_n) \label{delta}  \\ j_\theta(r) =
{\frac{e\hbar}{m_r}} {\frac{1}{r}}\sum_n \biggl[ \biggl(
l-{\frac{m}{2}}-{\frac{e}{\hbar c}} r A_\theta(r) \biggl)
|u_n(r)|^2 f_n \biggr. \nonumber \\ \biggl.
{}-\biggl(l+{\frac{m}{2}}+{\frac{e}{\hbar c}} r A_\theta(r)
\biggr)|v_n(r)|^2 (1 - f_n)\biggr], \label{current}
\end{eqnarray}
where $g$ is the coupling strength for the electron-electron attraction, $%
\omega_D$ is the Debye frequency, and $f_n\equiv f(\epsilon_n)$ is
the Fermi distribution function. The vector potential is given in
turn by the current density through the Maxwell equation $\nabla\times\nabla\times{\bf A}%
=(4\pi/c){\bf j}$. We first solve Eq.~(\ref{bdgeq2}) with initial
guesses for $|\Delta(r)|$ and $A_{\theta}(r)$ and recalculate them
from Eqs.~(\ref{delta}) and (\ref{current}), and repeat the
process until self-consistency is acquired. The local tunneling
density of states and the differential conductance
\cite{gygi91,tanaka93} can then be calculated by
\begin{eqnarray}  \label{ldos}
N(r,E)=\sum_{n} \bigl[|u_n(r)|^2 \delta(E-\epsilon_n) \bigr.
\nonumber\\ {}+ \bigl. |v_n(r)|^2 \delta(E+\epsilon_n) \bigr]
\\
{\frac{dI(r,V)}{dV}}\propto - \sum_{n}\bigl[|u_n(r)|^2f^%
\prime(\epsilon_n-eV) \bigl. \nonumber \\ {} +\bigl. |v_n(r)|^2
f^\prime(\epsilon_n+eV)\bigr]. \label{diffc}
\end{eqnarray}
Clearly, the differential tunneling conductance (\ref{diffc}) is a
direct probe of the local density of states, $N(r,E)$, provided
that the temperature is low enough. In the following we will
present some results for the experimentally observable
differential conductance at low temperatures, but may refer to it
as the local tunneling density of states (LDOS). We choose the
parameters corresponding to ${\rm NbSe}_2$: we take $m_r=2m_e$,
$E_F=37.3\,{\rm meV}$, $\hbar\omega_D=3.0\,{\rm meV}$, and set the
coupling strength so that the bulk gap $\Delta_0\equiv 1.12 \,{\rm
meV}$.

The results are shown for disk radius $R=500\,{\rm nm}$. We have
investigated the size range of $R=200-600\,{\rm nm}$ and have
found qualitatively same features in the LDOS.

\section{Results}
\label{sec:results}

\begin{figure}[tbp]

\includegraphics[width=8.0cm]{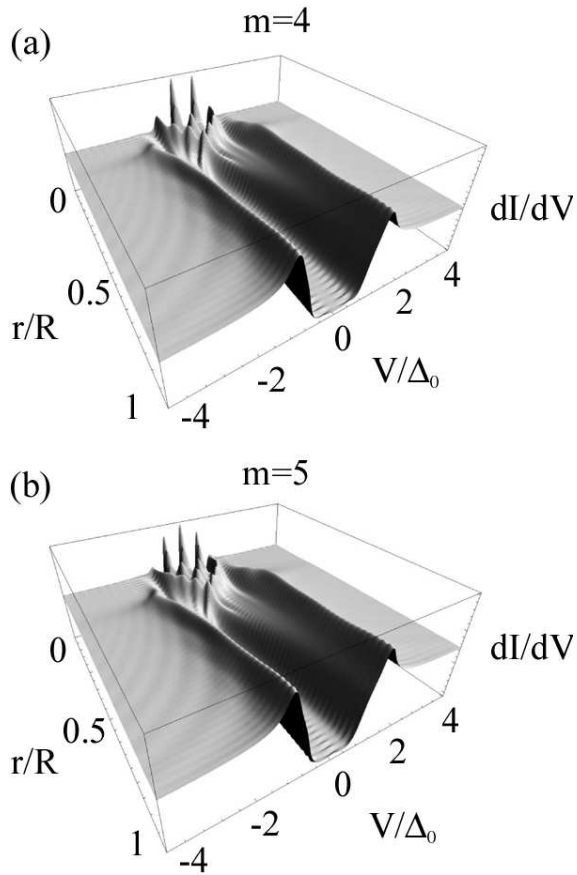}

\caption{Local tunneling conductance as a function of coordinate
$r$ and voltage $V$, for a giant vortex state with (a) $m=4$ and
(b) $m=5$ flux quanta, sustained in a superconducting disk with
radius $R=500\, {\rm nm}$ at temperature $T=1 \,K$.}

\label{fig:ldos}

\end{figure}

In Fig.~1 we show the differential conductance for a vortex with
(a) $m=4$ and (b) $m=5$ flux quanta, as a function of voltage $V$
and radial distance $r$ from the disk center, for temperature $T=1
K$ \footnote{For data to be shown for energies well above the gap
energy, fast $1/k_F$ oscillations have been removed by means of
Fourier transform. In actual observations these fast oscillations
will not be resolved.}. In both cases, prominent sharp peaks can
be seen near the vortex core and for low voltages -- four peaks in
the former and five in the latter. Generally speaking, the number
of low-bias conductance peaks corresponds to the winding number
$m$ of the order parameter, which gives rise to $m$ peaks near the
center \cite{tanaka93,virtanen99}. This feature is in accordance
with the index theorem established by Volovik \cite{volovik93} for
the Caroli-de Gennes-Matricon bound states in a vortex core.
According to this theorem, the quasiparticle spectrum of a vortex
with winding number $m$ has $m$ branches of bound states, which
cross zero energy as a function of angular momentum. These
quasiparticle branches also explain the evolution of the $m$ rows
of conductance peaks as one moves away from the core, as seen in
Fig.~1, with decreasing number of peaks one by one
\cite{tanaka93}. In contrast to the singly quantized case
\cite{shore}, one can see directly that as energy increases, more
states with higher angular momenta contribute to the density of
states.

We illustrate this in Fig.~2, where a spatial map of the LDOS is
taken for various fixed values of $V$. Clearly, with increasing
bias voltage the density of states is redistributed from the core
towards the sample boundaries. Fig.~1 also reveals the presence of
the so-called zero mode, i.e., a peak around zero energy at the
vortex core for odd $m$, and its absence for even $m$. The
existence of a zero mode is, quite generally, linked to a sign
change in the order parameter as a function of some generalized
coordinate \cite{janko}. The zero-bias peak is a signature of
bound states for quasiparticles trapped by the sign change. Here,
in the given gauge, the order parameter changes sign at the vortex
core when $m$ is odd, while it does not when $m$ is even.

The low-bias peaks and zero modes discussed above are general
characteristics of the LDOS associated with the winding number of
the order parameter. In addition to these, however, we have found
novel features in the LDOS that are unique to giant vortex states
in submicron disks. They are the oscillations seen above the gap
energy in Figs.~1 and 2(c) and more clearly in the contour plot of
Fig.~3. These oscillations are similar in origin to the so-called
Tomasch oscillations discovered in a superconductor-normal metal
junction \cite{tomasch65,tomasch66,mcmillan66} and reflect
``standing waves'' arising from the interference of quasiparticle
states. This interference effect is a direct consequence of strong
confinement experienced by the superconducting quasiparticles due
to the small system size.  We dedicate the remainder of this paper
to detailed discussions of this effect.

\begin{figure}[tbp]

\includegraphics[width=7.0cm]{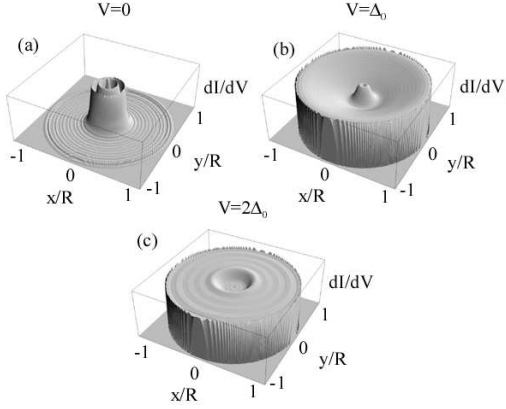}

\caption{Spatial map of the local tunneling conductance for the
entire disk in the giant vortex state of Fig.~1(b), for various
fixed voltages. It can be seen that the maxima in the local
density of states gradually shift towards the perimeter of the
disk as the voltage is increased.} \label{istvan}
\end{figure}

When a vortex holds multiple flux quanta $m$, the order parameter
vanishes around the center over a certain area -- the larger the
$m$ is, the larger the area \cite{virtanen99}. As the distance
from the center increases, the order parameter increases and
recovers to its bulk value eventually. In the case of ${\rm
NbSe}_2$, due to the short coherence length, the recovery happens
relatively quickly. This results in a well-defined superconducting
region with the constant order parameter $\Delta_0$ within the
disk.  At the disk boundaries, however, the order parameter is
forced to vanish and as a result, exhibits Friedel-like
oscillations around its bulk value near the surfaces.  These
oscillations have the largest amplitudes at the boundaries and
decay roughly over the coherence length scale.  Moreover, the
smaller the system size, the larger these amplitudes relative to
the bulk value. The quasiparticles confined in the disk experience
scattering by this large change in the order parameter at the
surfaces. An electron-like quasiparticle is reflected back as a
hole-like one and vice versa, and the Tomasch effect results from
the interference between the electron-like and hole-like states in
the superconducting region \cite{mcmillan66}.  The momenta of
electron-like and hole-like quasiparticles for energy $E$ are $k^{\pm} = {\frac%
{\sqrt{2 m_r}}{\hbar}}\sqrt{E_F \pm \Omega}\; \simeq\; k_F \pm {\frac%
{\Omega}{\hbar v_F}}$, respectively, where
$\Omega=\sqrt{E^2-|\Delta(r)|^2}$ with $|\Delta(r)|\simeq
\Delta_0$ and $k_F$ is the Fermi momentum. At a given distance $d$
from the surface, the LDOS oscillations in energy are determined
by \cite{mcmillan66} ${\frac{E_n }{\Delta_0}}\; \simeq\; \sqrt{1 +
n^2 \biggl({\frac{\pi \hbar v_F}{\Delta_0 d}}\biggr)^2}\;,$ where
$n$ is an integer, and ${\frac{\pi \hbar v_F}{\Delta_0}}\simeq
151.15\,{\rm nm}$ for ${\rm NbSe}_2$. Furthermore, the
interference can be seen in the LDOS also as a function of
coordinate (distance from the surface) for a given energy $E$. The
period of the oscillations in this case is given by
$\delta d\; \simeq \; {\frac{\pi \hbar v_F}{\Delta_0}} {\frac{1}{\sqrt{%
(E/\Delta_0)^2-1}}}\;.$  The oscillation periods obtained in our
numerical results, as seen in Fig.~3, are in quantitative
agreement with these analytical expectations.

\begin{figure}[tbp]

\includegraphics[width=7.0cm]{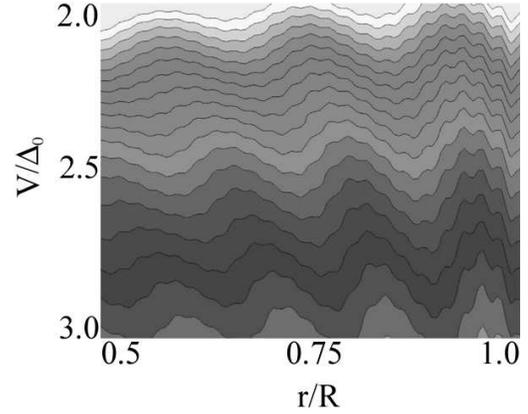}

\caption{Contour plot of the LDOS in Fig.~1(b) in the
superconducting region, where the Tomasch density of states
oscillations occur.} \label{contours}
\end{figure}

\begin{figure}[tbp]

\includegraphics[width=7.0cm]{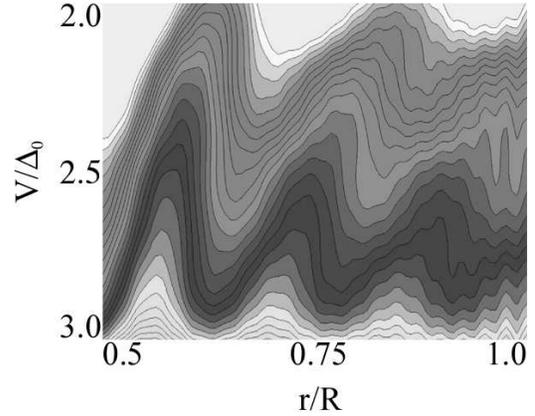}

\caption{Same as Fig.~\ref{contours}, but for the corresponding
model calculation.} \label{contourm}
\end{figure}

In a superconductor with short coherence length, if the winding
number $m$ and, consequently, the ``normal core'' area is very
large, the Tomasch effect may arise also from the vortex core, as
in a normal-superconductor junction. For submicron ${\rm NbSe}_2$
disks with $m$ up to five, however, we have found that the LDOS is
dominated by the Tomasch oscillations coming from the surfaces.
Indeed, we have confirmed the LDOS oscillations characteristic of
the Tomasch effect in terms of model calculations in one and two
dimensions, where the BdG equations are solved without iteration
with a step-function order parameter: $\Delta(r)=0 (r<R/2);
\Delta_0 (r>R/2)$. In this case the Tomasch effect coming from the
normal-superconductor interface governs the LDOS structure, so
that the oscillation period in energy (see $E_n$ above) becomes
larger as one approaches the interface (i.e., here $d$ is the
distance from the interface). Apart from this and enhanced
amplitudes due to a larger change in the order parameter, the LDOS
shows the same qualitative features as seen above (compare Figs.~3
and 4).

\section{Conclusions}
\label{sec:conclusions}

We have presented detailed, self-consistent calculations of the
microscopic electronic structure of giant vortex states. We
believe that the most direct experimental evidence for the
existence of giant vortices can be provided by STM measurements of
the local density of states in sub-micron superconductors capable
of sustaining such vortex configurations. We have provided a
spatial map of the tunneling density of states for multi-quantum
giant vortex states, and have identified several signatures that
can be used to identify them with STM. We have found that under
extreme confinement the quantum interference arises among
quasiparticle states and leads to experimentally observable
Tomasch oscillations in the local density of states.

\section{Acknowledgments}
We would like to thank  Prof.~A. A. Abrikosov, Dr.~G. W. Crabtree,
Dr.~W. K. Kwok, Prof.~F. Marsiglio, and Dr.~O. Tchernyshyov for
enlightening discussions. One of us (B. J.) would like to thank
Prof.~W. Tomasch for comments and discussions on the presented
results. This research was supported by U.S. DOE, Office of
Science, under contract No. W-31-109-ENG-38, and by the Natural
Sciences and Engineering Research Council of Canada.

Reprint requests should be addressed to B. J., e-mail:
bjanko@nd.edu.


\end{document}